\documentclass[prb,twocolumn,showpacs,preprintnumbers,amsmath,aps]{revtex4}
\usepackage{graphicx,hyperref}
\usepackage{xcolor}
\usepackage{bm}
\usepackage{subfigure}

\newcommand\DR{\text{DR}}

\def\nbC{{\mathchoice {\setbox0=\hbox{$\displaystyle\rm C$}%
\hbox{\hbox to0pt{\kern0.4\wd0\vrule height0.9\ht0\hss}\box0}}
{\setbox0=\hbox{$\textstyle\rm C$}\hbox{\hbox
to0pt{\kern0.4\wd0\vrule height0.9\ht0\hss}\box0}}
{\setbox0=\hbox{$\scriptstyle\rm C$}\hbox{\hbox
to0pt{\kern0.4\wd0\vrule height0.9\ht0\hss}\box0}}
{\setbox0=\hbox{$\scriptscriptstyle\rm C$}\hbox{\hbox
to0pt{\kern0.4\wd0\vrule height0.9\ht0\hss}\box0}}}}
\def\nbQ{{\mathchoice {\setbox0=\hbox{$\displaystyle\rm
Q$}\hbox{\raise
0.15\ht0\hbox to0pt{\kern0.4\wd0\vrule height0.8\ht0\hss}\box0}}
{\setbox0=\hbox{$\textstyle\rm Q$}\hbox{\raise
0.15\ht0\hbox to0pt{\kern0.4\wd0\vrule height0.8\ht0\hss}\box0}}
{\setbox0=\hbox{$\scriptstyle\rm Q$}\hbox{\raise
0.15\ht0\hbox to0pt{\kern0.4\wd0\vrule height0.7\ht0\hss}\box0}}
{\setbox0=\hbox{$\scriptscriptstyle\rm Q$}\hbox{\raise
0.15\ht0\hbox to0pt{\kern0.4\wd0\vrule height0.7\ht0\hss}\box0}}}}
\def\nbT{{\mathchoice {\setbox0=\hbox{$\displaystyle\rm
T$}\hbox{\hbox to0pt{\kern0.3\wd0\vrule height0.9\ht0\hss}\box0}}
{\setbox0=\hbox{$\textstyle\rm T$}\hbox{\hbox
to0pt{\kern0.3\wd0\vrule height0.9\ht0\hss}\box0}}
{\setbox0=\hbox{$\scriptstyle\rm T$}\hbox{\hbox
to0pt{\kern0.3\wd0\vrule height0.9\ht0\hss}\box0}}
{\setbox0=\hbox{$\scriptscriptstyle\rm T$}\hbox{\hbox
to0pt{\kern0.3\wd0\vrule height0.9\ht0\hss}\box0}}}}
\def\nbS{{\mathchoice
{\setbox0=\hbox{$\displaystyle     \rm S$}\hbox{\raise0.5\ht0%
\hbox to0pt{\kern0.35\wd0\vrule height0.45\ht0\hss}\hbox
to0pt{\kern0.55\wd0\vrule height0.5\ht0\hss}\box0}}
{\setbox0=\hbox{$\textstyle        \rm S$}\hbox{\raise0.5\ht0%
\hbox to0pt{\kern0.35\wd0\vrule height0.45\ht0\hss}\hbox
to0pt{\kern0.55\wd0\vrule height0.5\ht0\hss}\box0}}
{\setbox0=\hbox{$\scriptstyle      \rm S$}\hbox{\raise0.5\ht0%
\hboxto0pt{\kern0.35\wd0\vrule height0.45\ht0\hss}\raise0.05\ht0%
\hbox to0pt{\kern0.5\wd0\vrule height0.45\ht0\hss}\box0}}
{\setbox0=\hbox{$\scriptscriptstyle\rm S$}\hbox{\raise0.5\ht0%
\hboxto0pt{\kern0.4\wd0\vrule height0.45\ht0\hss}\raise0.05\ht0%
\hbox to0pt{\kern0.55\wd0\vrule height0.45\ht0\hss}\box0}}}}
\def\nbZ{{\mathchoice {\hbox{$\sf\textstyle Z\kern-0.4em Z$}}
{\hbox{$\sf\textstyle Z\kern-0.4em Z$}}
{\hbox{$\sf\scriptstyle Z\kern-0.3em Z$}}
{\hbox{$\sf\scriptscriptstyle Z\kern-0.2em Z$}}}}

\begin{document}

\title{Dimensional reduction breakdown and correction to scaling in the random-field Ising model}

\author{Ivan Balog} \email{balog@ifs.hr}
\affiliation{Institute of Physics, P.O.Box 304, Bijeni\v{c}ka cesta 46, HR-10001 Zagreb, Croatia}

\author{Gilles Tarjus} \email{tarjus@lptmc.jussieu.fr}
\affiliation{LPTMC, CNRS-UMR 7600, Sorbonne Universit\'e,
bo\^ite 121, 4 Pl. Jussieu, 75252 Paris cedex 05, France}

\author{Matthieu Tissier} \email{tissier@lptmc.jussieu.fr}
\affiliation{LPTMC, CNRS-UMR 7600, Sorbonne Universit\'e,
bo\^ite 121, 4 Pl. Jussieu, 75252 Paris cedex 05, France}

\date{\today}

\begin{abstract}
We provide a theoretical analysis by means of the nonperturbative functional renormalization group (NP-FRG) of the corrections to scaling in the critical behavior of the random-field Ising model (RFIM) near the dimension $d_{DR}\approx 5.1$ that separates a region where the renormalized theory at the fixed point is supersymmetric and critical scaling satisfies the $d\to d-2$ dimensional reduction property ($d>d_{DR}$) from a region where both supersymmetry and dimensional reduction break down at criticality ($d<d_{DR}$). We show that the NP-FRG results are in very good agreement with recent large-scale lattice simulations of the RFIM in $d=5$ and we detail the consequences for the leading correction-to-scaling exponent of the peculiar boundary-layer mechanism by which the dimensional-reduction fixed point disappears and the dimensional-reduction-broken fixed point emerges in $d_{DR}$. 
\end{abstract}


\maketitle

\section{Introduction}

One of the puzzles raised by the critical behavior of the random-field Ising model is the way the underlying supersymmetry (SUSY) unveiled by Paris and Sourlas\cite{parisi79} and the associated dimensional-reduction property (according to which the critical behavior of the RFIM in dimension $d$ is the same as the critical behavior of the pure Ising model in $d-2$) break down as a function of decreasing dimension. Dimensional reduction has rigorously been shown to be wrong in dimensions $d=3$ and $d=2$\cite{imbrie84, bricmont87,aizenman89} but is valid at the upper critical dimension $d=6$ (although it only has a weak meaning there). By using a nonperturbative functional renormalization group (NP-FRG) approach,\cite{tarjus04,tissier06,tissier11,FPbalog,tarjus20} we showed that the solution of the puzzle is the existence of a nontrivial critical dimension $d_{DR}\approx 5.1$ above which dimensional reduction applies for the scaling behavior of the critical point while SUSY is obeyed at the associated zero-temperature fixed point. Below $d_{DR}$ both SUSY and dimensional reduction are broken. Our predictions have since been supported by large-scale state-of-the-art lattice simulations in dimensions up to $d=5$.\cite{fytas17,fytas18,fytas19} 

One point however remains puzzling, which has been recently raised by Fytas and coworkers.\cite{fytas19} If the critical dimension is indeed near $5.1$, the corrections to scaling around the critical point of the RFIM in $d=5$ should keep track of an eigenvalue that vanishes for $d=d_{DR}$ and is then parametrically small in $d_{DR}-d\approx0.1$. Yet, this does not seem to be seen in the simulation study of the $5$-d RFIM in which one instead finds a correction-to-scaling exponent $\omega =0.66(+15/-13)$.\cite{fytas17,fytas19} As argued in Ref.~[\onlinecite{fytas19}] the correction-to-scaling exponent may then be the ``smoking gun'' to settle whether our scenario of a critical dimension in $d_{DR}\approx 5.1$ is valid or whether violation of supersymmetry (and consequently of dimensional reduction) is always present below the upper critical dimension $d=6$ but is exponentially small in $1/(6-d)$.

In this paper we revisit the critical behavior of the RFIM predicted by the NP-FRG in the vicinity of the critical dimension $d_{DR}$. We show that the peculiar way by which the new fixed point emerges from the collapse of a pair of dimensional-reduction fixed points (one stable and one unstable) in $d=d_{DR}$\cite{FPbalog} allows one to escape the curse of the small correction-to-scaling exponent $\omega$. It indeed leads to a rapid increase of $\omega$ below $d_{DR}$, in the form of a square-root singularity, which results in a predicted value of the exponent $\omega$ that is compatible with the numerical finding of Refs.~[\onlinecite{fytas17,fytas19}]. Furthermore we argue that in such a scenario of disappearance and emergence of fixed points, it would in any case be extremely difficult to detect the true asymptotic critical behavior if one happens to get very close to the critical dimension $d_{DR}$ because of the existence of a ``Larkin length'' that diverges extremely rapidly, as an exponential of $1/\sqrt{d_{DR}-d}$, as one approaches $d_{DR}$ from below.

The rest of the paper is organized as follows. In the following section, Sec.~\ref{sec:model}, we recall the definition of the model and of its field-theoretical version and we give a brief recap of the NP-FRG approach. In Sec.~\ref{sec:scenario}, we discuss the scenario found within the NP-FRG for the appearance and disappearance of the critical fixed points near the critical dimension $d_{DR}$ and its consequences for the correction-to-scaling exponent $\omega$ and the Larkin length. In Sec.~\ref{sec:d=5} we present the results obtained within our NP-FRG approach for the critical exponents in $d=5$ and compare them with recent lattice simulations. We next study in Sec.~\ref{sec:Larkin_length} the appearance of a Larkin length along the NP-FRG flow for $d<d_{DR}$ and we discuss its meaning and implications. In Sec.~\ref{sec:corrections} we consider the corrections to scaling in the immediate vicinity of $d_{DR}$ and we discuss separately the region below and above $d_{DR}$. Finally we provide some discussion and concluding remarks in Sec.~\ref{sec:conclusion}. Additional details are given in several  appendices.

\section{The RFIM}
\label{sec:model}

The RFIM is defined on a $d$-dimensional hypercubic lattice by the following Hamiltonian\cite{imry-ma75,nattermann98}
\begin{equation}
\label{eq_Ham_dis}
H[S;h]=  _J\sum_{<ij>}S_iS_j -\sum_i h_i S_i
\end{equation}
where $S_i=\pm1$ and the $h_i$'s are independently distributed quenched random fields. To investigate the critical behavior of the model by the RG it is convenient to rather start with the field-theoretical version of the RFIM. The associated ``bare action'' is then
\begin{equation}
\label{eq_action_dis}
S[\varphi;h]=  S_B[\varphi]-\int_{x} h(x) \varphi(x)
\end{equation}
with
\begin{equation}
\label{eq_bare-action}
S_B[\varphi]= \int_{x}\bigg\{\frac{1}{2}[\partial \varphi(x)]^2+ \frac{r}{2} \varphi(x)^2 + \frac{u}{4!} \varphi(x)^4 \bigg\},
\end{equation}
where $ \int_{x} \equiv \int d^d x$ and $h(x)$ is a random ``source'' (a random magnetic field). This quenched random field $h(x)$ is taken with a Gaussian distribution characterized by a zero mean and a variance $\overline{h(x)h(y)}= \Delta_B \delta^{(d)}( x-y)$ (and similarly for the lattice version).

We are interested in the equilibrium thermodynamic properties of the model and, more precisely, in the critical behavior. The critical point is controlled by a zero-temperature fixed point and its characteristics can be studied directly at zero temperature by focusing on the properties of the ground state which   is what we do in this paper. This is in this zero-temperature setting that SUSY emerges in the theory.\cite{parisi79}

In the NP-FRG approach of the RFIM,\cite{tarjus04,tissier06,tissier11,tarjus20} the central quantities are the cumulants of the effective action at a running infrared scale $k$, which generate all the $1$-particle irreducible correlation functions. One focuses in particular on the effective potential, from which the thermodynamics is obtained, on the field renormalization, from which one derives the anomalous dimensions, and on the cumulants of the renormalized random field, which are field derivatives of the cumulants of the effective action. In this framework, the breaking of SUSY at the fixed point and the breakdown of the dimensional-reduction predictions for scaling around the critical point are related to the appearance of a singular behavior, in the form of linear cusps, in the functional dependence of the dimensionless cumulants of the renormalized random field when two field arguments become equal, {\it e.g.}, for the second cumulant $\delta_*$ and uniform field configurations,
\begin{equation}
\label{eq_cusp}
\delta_*(\varphi-\delta\varphi,\varphi+\delta\varphi)-\delta_*(\varphi,\varphi) \propto \vert \delta\varphi\vert\,,\;{\rm when}\;  \delta\varphi \to 0\,.
\end{equation}
The cusps generate a breakdown of the Ward-Takahashi identities associated with SUSY, with ensuing spontaneous SUSY breaking and dimensional-reduction breakdown.\cite{tissier11} These cusps have been shown to be intimately related to the presence of ``static avalanches'' or ``shocks''\cite{aval_REMM} in the variation of the ground state of the RFIM with the external source (magnetic field). Avalanches on all scales are always present in the critical behavior of the RFIM at zero temperature but in the region where dimensional reduction and SUSY are valid, the effect of these avalanches is too weak to influence the critical physics.\cite{tarjus13} (Note that this subdominant behavior and the associated irrelevant eigenvalue around the cuspless  fixed point can be accessed through a perturbative but functional RG treatment of the RFIM in an expansion in $\epsilon=6-d$ around the upper critical dimension: see Ref.~[\onlinecite{tissier_pertFRG}].) It is only below a critical dimension $d_{DR}$ that the avalanches and the resulting cusps appear at the level of the fixed point.\cite{tarjus13} 

A brief recap on our NP-FRG approach is given in  Appendix~\ref{app:NP-FRG} (see also Ref. [\onlinecite{tarjus20}]).

\section{Appearance and disappearance of the fixed points near the critical dimension $d_{DR}$}
\label{sec:scenario}

The way the fixed point associated with SUSY and dimension reduction disappears when $d$ decreases and a new fixed point emerges has been studied in a previous paper.\cite{FPbalog} What is found within the NP-FRG is that the SUSY/dimensional-reduction fixed point that controls the critical behavior of the RFIM below the upper dimension $d=6$ annihilates with another (unstable) SUSY/dimensional-reduction fixed point when $d=d_{DR}$. Then emerges from these collapsed fixed points a ``cuspy'' fixed point at which both SUSY and dimensional reduction are broken. This fixed point appears below $d_{DR}$ through an unusual {\it boundary-layer mechanism} that can only be captured through a functional approach.\cite{footnote_BL}

One of the consequences of this boundary-layer mechanism is that the dependence of the lowest positive (irrelevant) eigenvalues of the stability matrix at the critical fixed point have a singular square-root behavior around $d_{DR}$. If we denote by $\lambda$ the lowest eigenvalue associated with a ``cuspy'' perturbation around the fixed point (above $d_{DR}$ the fixed point itself is ``cuspless'' whereas it is ``cuspy'' below $d_{DR}$),  we find that $\lambda$ approaches a small but nonzero value with a {\it square-root singularity} when $d\to d_{DR}^+$, has a {\it discontinuity} in $d_{DR}$, and below $d_{DR}$ increases from $0$ with another square-root behavior:
\begin{equation}
\begin{aligned}
\label{eq_lambda}
\lambda(d)&\approx \lambda_\ddagger + c_+ \sqrt{d-d_{DR}}\,,\;\; {\rm for} \; d>d_{DR}\,,\\&
\approx c_- \sqrt{d_{DR}-d}\,,\; \;\;\;\;\;\;\;\;\;\,{\rm for} \;  d<d_{DR}\,,
\end{aligned}
\end{equation}
with $\lambda_\ddagger>0$ and the constants $c_{\pm}$ positive and of O($1$). Above $d_{DR}$, there are additional irrelevant eigenvalues which display a square-root, $\sqrt{d-d_{DR}}$, dependence and are associated with cuspless (but possibly showing weaker singularities in the form of  ``subcusps'') perturbations in the $2$-replica sector. There is also the correction-to-scaling exponent $\omega_{DR}$ obtained by dimensional reduction (and therefore characterizing the $1$-replica sector). All of this will be again discussed in Sec.~\ref{sec:corrections}.

\begin{figure}[tbp]
\includegraphics[width=.9 \linewidth]{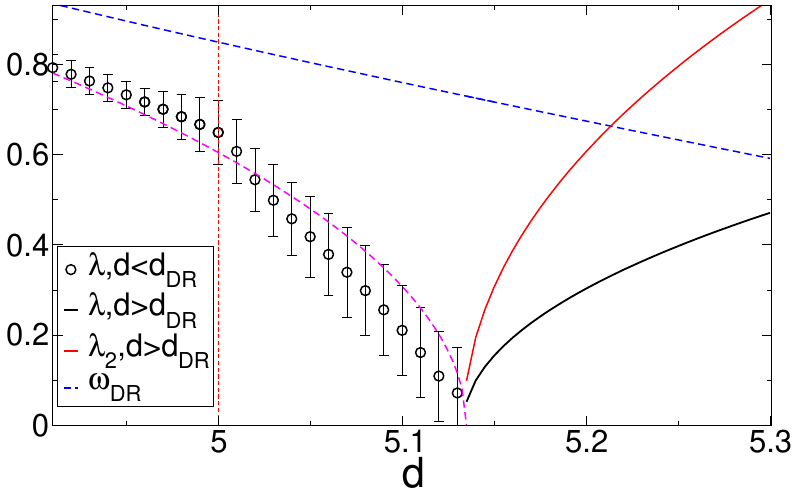}
\caption{Lowest irrelevant eigenvalues of the stability matrix of the RFIM fixed point numerically obtained from our NP-FRG approach in the vicinity of the critical dimension $d_{DR} \approx 5.13$.   The lowest eigenvalue $\lambda$ is discontinuous in $d_{DR}$ and has square-root singularities on approaching $d_{DR}$ from both sides (dashed and full lines). (As the numerical resolution of the FRG equations is done by discretizing the field arguments on a grid,\cite{tissier11,FPbalog} it is very difficult to properly approach $d_{DR}$ from below and the curve there is just a guide for the eye.) The eigenvalue denoted by $\omega_{DR}$ is the correction-to-scaling exponent obtained by dimensional reduction for the pure $\phi^4$ theory in two dimensions less (it vanishes as $\epsilon=6-d$ near $d=6$). This is a true eigenvalue of the RFIM fixed point above $d_{DR}$ but not below where dimensional reduction is broken.  Above $d_{DR}$ the eigenvalue $\lambda$ is associated with a cuspy perturbation around the cuspless fixed point. The red curve denoted $\lambda_2$ is associated with a cuspless perturbation in the $2$-replica component of the second cumulant of the renormalized random field (see the main text); it also follows a square-root dependence when approaching $d_{DR}$.}
\label{Fig_eigenvalues}
\end{figure}

This pattern of singular behavior is hard to analytically extract from the full NP-FRG equations for the RFIM. (It is on the other hand easily obtained in a toy model having a similar but simpler structure which we introduced in Ref.~[\onlinecite{FPbalog}] and is discussed in Appendix~\ref{app:toy}.) Numerical evidence from the NP-FRG is shown in Fig.~\ref{Fig_eigenvalues}, where we plot the three lowest eigenvalues as a function of $d$ in the region near $d_{DR}$. The square-root behavior of the lowest irrelevant eigenvalue, which is reasonably well followed by the numerical solution (although the limit $d \to d_{DR}^-$ is very hard to access due to the boundary-layer structure),  implies a rapid increase of this eigenvalue below $d_{DR}\approx 5.13$, at variance with the conventional behavior of a crossover exponent near the dimension at which there is an exchange of stability between two fixed points.

\begin{figure}[tbp]
\centering
\includegraphics[width=.9 \linewidth]{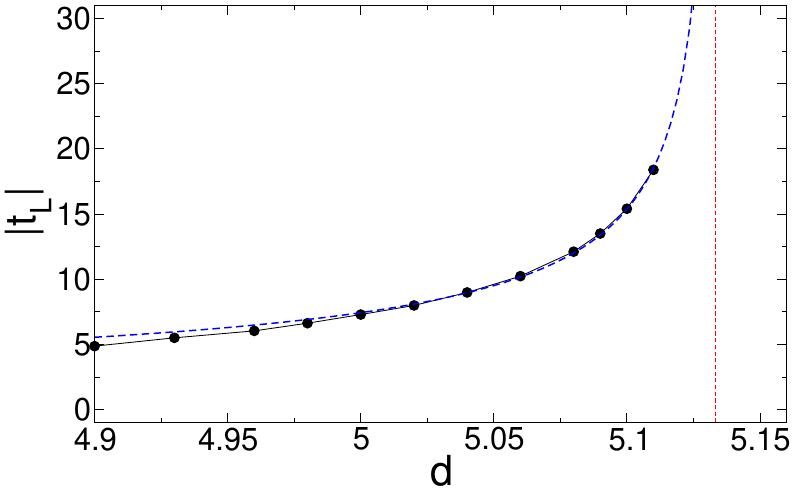}
\includegraphics[width=.9 \linewidth]{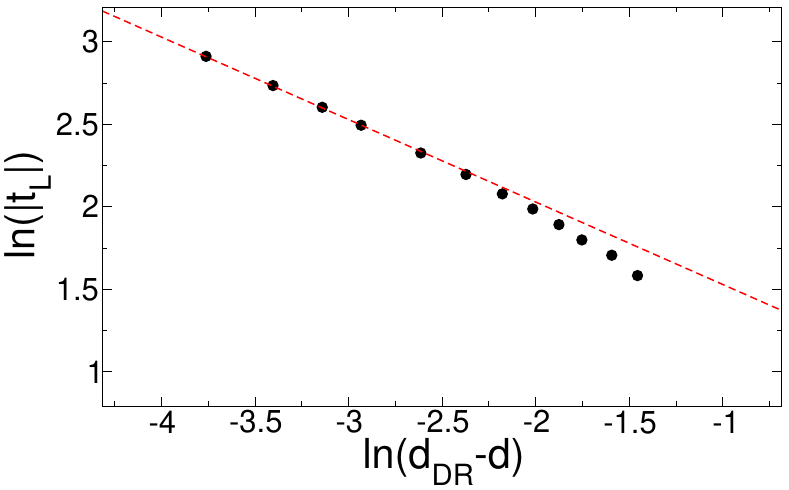}
\caption{Dimensional dependence of the Larkin length $\ell_L(d)$ for $d<d_{DR} \approx 5.13$ obtained from the NP-FRG solution of the RFIM at criticality. The flow starts with a cuspless initial condition for the second cumulant of the random field, 
$\delta_k(\varphi_1,\varphi_2)$. The cuspless solution is valid down to a value of the IR cutoff $k$ (or RG time $t=\ln(k/\Lambda)$ where $\Lambda$ is the UV scale) that can be easily found numerically from the divergence of the second derivative in the direction $\varphi_1-\varphi_2$ evaluated for equal replica fields, $\delta_{k,2}(\varphi)$: This is our definition of the Larkin length. Top panel: $\vert t_L\vert=\ln ( \ell_L \Lambda)$ versus $d$ for $d\leq d_{DR}=5.13325$ (as determined after optimizing the regulators); the dashed blue curve is the $1/\sqrt{d_{DR}-d}$ behavior. Bottom panel: log-log plot of $\vert t_L\vert$ versus $(d_{DR}-d)$; the dashed red line has a slope of $-1/2$.}
\label{Fig_Larkin_length}
\end{figure}

Another specific feature of the boundary-layer mechanism for the emergence of the cuspy fixed point below $d_{DR}$ is the existence of a large ``Larkin length'', $\ell_L$, which diverges when $d \to d_{DR}^-$. The notion of a Larkin length has been introduced in the context of elastic manifolds pinned by a random potential.\cite{larkin70,chauve_creep,chauve04} There, it marks the crossover between (small) lengths over which the behavior of the manifold is controlled by its elastic properties and (large) lengths over which the pinning potential dominates and induces a multiplicity of metastable states, avalanches between states,  and a linear cusp in the functional dependence of the cumulants of the renormalized random force.\cite{larkin70,chauve_creep,chauve04} In the case of the RFIM the physical nature of the Larkin length is more elusive. It can nonetheless be operationally obtained as the lenghscale along the functional RG flow at which a cusp first appears in the cumulants of the renormalized random field when starting from a cuspless initial condition at the UV (microscopic) scale (see Appendix~\ref{app:NP-FRG} for more details). The so-defined Larkin length diverges exponentially rapidly when approaching the critical dimension $d_\DR$ from below,
\begin{equation}
\label{eq_Larkin_length}
\ln \ell_L(d) \sim \frac 1{\sqrt{d_{DR}-d}} \,,
\end{equation}
where the square-root dependence is analytically found in the toy model already mentioned (see Appendix~\ref{app:toy}) and is also compatible with the numerical solution of the NP-FRG of the RFIM: see Fig.~\ref{Fig_Larkin_length}. There are prefactors in the above relation which depend on the microscopic details of the theory (and the Larkin length itself is expressed in terms of some microscopic length), but in any case the length becomes extremely large in the immediate vicinity of the critical dimension. A more extended discussion of the Larkin length in the RFIM will be given in Sec.~\ref{sec:Larkin_length}.

\section{NP-FRG results in $d=5$}
\label{sec:d=5}

As already mentioned, our FRG approach is nonperturbative but approximate. We have developed a truncation scheme of the running effective action that respects the symmetries and supersymmetries of the RFIM. Since the resulting functional flow equations require a numerical resolution, the best level of truncation of the NP-FRG that we have at present achieved consists in truncating the cumulant expansion after the second cumulant and truncating the expansion in spatial derivatives of the field after the second order:\cite{tarjus04,tissier11,FPbalog} see also Appendix~\ref{app:NP-FRG}. The results for the critical exponents are in very good agreement with the best estimates from simulations in all dimensions $d$.\cite{tissier11,tarjus_exponents,tarjus20} With this truncation we find that SUSY and dimensional reduction are valid at the (cuspless) fixed point above $d_{DR}\approx 5.13$. 

More specifically in $d=5$ we have obtained the following results: The anomalous dimensions are found to be $\eta\approx 0.044$, $\bar\eta \approx 0.048$, whereas the correlation-length exponent is  $\nu\approx 0.627$.\cite{tissier11,FPbalog} The anomalous dimensions very slightly break the SUSY related relation $\eta=\bar\eta$. The exponents also slightly deviate from the dimensional-reduction property. (In the same approximation we indeed obtain for the pure $3$-$d$ Ising model $\eta\approx0.045$ and $\nu\approx 0.628$.) Finally, we also find that the correction-to-scaling exponent is $\omega=\lambda\approx 0.65$. 

These values are in excellent agreement with those reported in Ref.~[\onlinecite{fytas17}] from state-of-the-art computer simulations of the RFIM at zero temperature: $\eta=0.052$, $\bar\eta=0.058$, $\nu=0.626$, which also point to a very slight violation of dimensional reduction. Notably, the correction-to-scaling exponent is compatible with the simulation result $\omega=0.66(+15/-13)$.\cite{footnote_omega}

One, of course, has to be careful before drawing definite conclusions. Both the NP-FRG and the simulation results are affected by uncertainties. They result from the truncation of the effective average action in the former and to the limited system sizes and limited number of samples in the latter. (Due to the uncertainties in the simulations, the authors of Ref.~[\onlinecite{fytas17}] and of the more recent work in Ref.~[\onlinecite{fytas19}] for instance comment that their results are also compatible with dimensional reduction\cite{fytas17} and with SUSY.\cite{fytas19}) In the case of the NP-FRG result it would be interesting  to be able to assess the uncertainty associated with the truncation. While expected to be small for most critical exponents, as shown for a variety of models studied by the FRG, this uncertainty may be quite significant in the case of the correction-to-scaling exponent because of the uncertainty on the precise value of the critical dimension $d_{DR}$. To give a rough estimate, let us assume that the uncertainty on the determination of $d_{DR}$ is, say, $\pm 0.05$, {\it i.e.},  $5.08\leq d_{DR}\leq 5.18$, and that we can use the square-root expression in Eq. (\ref{eq_lambda}) with the same parameter $c_-$. We then find that $\omega$ would vary from $0.5$ at the lower edge to $0.7$ at the upper edge. This is still compatible with the results from simulation. Of course it is highly desirable to have an actual prediction for the uncertainty on $d_{DR}$ and we are presently working on this (difficult) issue.

In any case, the overall agreement between NP-FRG and simulation results appears as a strong support for our scenario of dimensional-reduction and SUSY breaking below $d_{DR}\approx 5.1$.

\begin{figure}[tbp]
\centering
\includegraphics[width=.9 \linewidth]{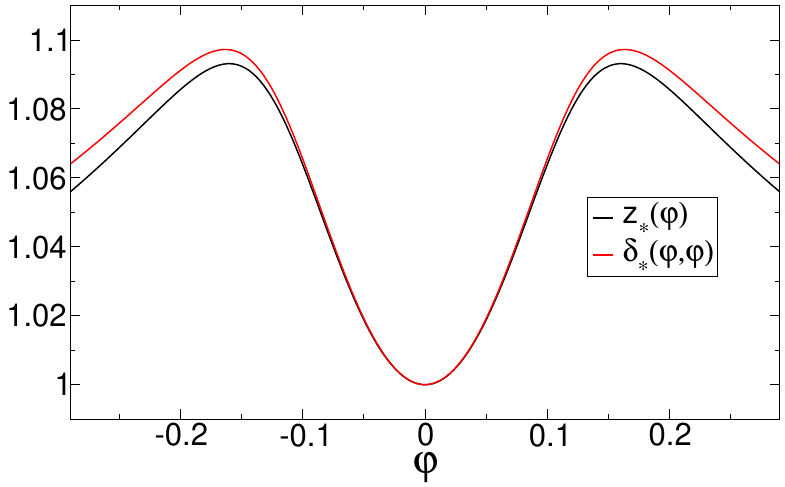}
\caption{Test of the SUSY Ward identity at the NP-FRG fixed point in $d=5$. SUSY implies that the second cumulant of the renormalized random field $\delta_*(\varphi,\varphi)$ is equal to the field renormalization function $z_*(\varphi)$;\cite{tissier11} The observed violation is by less than $1\%$, as found for a related Ward identity in a lattice simulation\cite{fytas19}.}
\label{Fig_SUSY}
\end{figure}

Finally, the NP-FRG also provides fixed-point functions in addition to critical exponents. The existence of a violation of SUSY can therefore be tested at the NP-FRG fixed point through the SUSY-induced Ward identity which predicts a relation between the second cumulant of the renormalized random field for equal arguments, $\delta_*(\varphi,\varphi)$, and the field renormalization function, $z_*(\varphi)$, which is the derivative with respect to $q^2$ of the $q$-dependent $2$-point vertex obtained from the first cumulant of the effective action:\cite{tissier11} 
\begin{equation}
\label{eq_SUSYward}
\delta_*(\varphi,\varphi)=z_*(\varphi).
\end{equation}
With the truncation that we use, this identity is exactly satisfied above $d_{DR}$. The comparison in $d=5$ between the left and right members of the above equation is shown in Fig.~\ref{Fig_SUSY}. One can see that {\it the violation of the SUSY Ward identity is indeed extremely small} (less than $1\%$), as also found in Ref.~[\onlinecite{fytas19}] for a related Ward identity.

\section{The Larkin length when $d\to d_{DR}^-$}
\label{sec:Larkin_length}

We now come back to the issue of the physical interpretation of the Larkin length $\ell_L$ in the vicinity, but below, $d_{DR}$ and on its consequences. We would like to better understand the meaning of the Larkin scale as it appears in the NP-FRG flow and relate it to some characteristic length in a microscopic realization of the RFIM in a $d$-dimensional lattice at zero temperature. In principle, if one wishes to make a direct comparison between nonuniversal quantities obtained from the field-theoretical NP-FRG on the one hand and from lattice simulations on the other, and not only focus on universal predictions, one should consider the hard-spin lattice version of the NP-FRG, as developed for instance in Ref.~[\onlinecite{machado10}]. In the latter approach, the initial condition of the NP-FRG flow is provided by the limit of decoupled sites in the original system. In the present case this limit implies computing the effective action and all the cumulants of the renormalized random field for a single-site RFIM, which can be done essentially exactly. 

One subtlety that appears in the case of the RFIM is that two distinct IR cutoff functions are generically introduced to derive exact FRG equations (see Appendix~\ref{app:NP-FRG}):\cite{tissier11} One of them, which we called $\widehat R_k$ in Ref.~[\onlinecite{tissier11}], restricts the fluctuations of the fundamental field (the local magnetization in the language of magnetic systems) with momenta less than the running scale $k$ and is furthermore chosen to enforce site decoupling in the lattice NP-FRG at a UV wavevector $k_{UV}=2\pi/a$ where $a$ is the lattice spacing; the other one, which we called $\widetilde R_k$ in Ref.~[\onlinecite{tissier11}], restricts the fluctuations associated with the random source (or field) and reduces the variance of the latter. In the continuum description, if one wishes to avoid an {\it explicit} breaking of the SUSY Ward identity, one must impose a relation between the two IR cutoff functions. In momentum space, the relation reads\cite{tissier11}
\begin{equation}
\label{eq_SUSYregulator}
\widetilde R_k(q^2)=-\frac{\Delta_k}{Z_k}\frac{\partial}{\partial q^2}\widehat R_k(q^2)\,,
\end{equation}
where $q^2$ is the squared momentum, $\Delta_k$ the (squared) strength of the renormalized random field and $Z_k$ the field renormalization constant (see Appendix~\ref{app:NP-FRG}). However, SUSY being broken at the fixed point in $d<d_{DR}$, there is some leeway in choosing the IR cutoff function $\widetilde R_k$ in this region as the same long-distance physics should be obtained when $k\to 0$ irrespective of the (reasonable) choices of IR regulators (at least in the exact FRG). 

For a given choice of $\widehat R_k$ there is therefore a spectrum of possible $\widetilde R_k$'s, from one that mimics the SUSY-induced relation and exactly cancels the bare variance of the random field at the UV scale to others that lead to only partial cancellation. Although the initial conditions of the FRG flow are different for each $\widetilde R_k$, the very same critical physics is attained in the limit where the IR cutoff $k$ goes to zero.\cite{footnote_fixed-points} Of course, when approximations are introduced, this is no longer exactly true and, {\it e.g.}, the critical exponents obtained through the different choices of $\widetilde R_k$ may not exactly coincide.

Without delving more into this question, we just note that the cumulants of the effective random field obtained from the solution of the single-site RFIM display a cusp in their functional field dependence for choices of $\widetilde R_k$ that do not exactly cancel the bare variance of the random field at the UV cutoff $k_{UV}$.\cite{tarjus20} One may therefore start the NP-FRG flow with or without a cusp in the cumulants. Only in the latter case will a crisp Larkin scale be observed along the flow; it will be replaced by a crossover in the former cases.

In the present work we do not attempt to study the full-blown lattice NP-FRG for the RFIM. We defer this to a future study. We rather use a proxy by considering the continuum field-theoretical NP-FRG, as previously investigated, in two different settings. In the first setting, we consider cuspless initial conditions and FRG flow equations obtained from the choice in Eq. (\ref{eq_SUSYregulator}). In the second one, we consider cuspy initial conditions and FRG flow equations derived with $\widetilde R_k(q^2)=-\alpha (\Delta_k/Z_k)(\partial/\partial q^2)\widehat R_k(q^2)$ with $\alpha <1$: for illustration we choose $\alpha=0.6$. (For completeness we also consider a case with $\alpha=1$ but with cuspy initial conditions.) In all of the cases, the initial condition for the first cumulant is a $\varphi^4$ theory as in Eq. (\ref{eq_bare-action}), with the bare potential rewritten as
\begin{equation}
\label{eq_phi4_UV}
u_\Lambda(\varphi)=\frac{g_{\Lambda}}{4!}(\varphi^2-\kappa_\Lambda)^2
\end{equation}
where we keep $g_\Lambda$, the $\varphi^4$ coupling constant, fixed and we fine-tune, for each choice of the initial cusp amplitude, the  location of the nontrivial minimum of the potential, $\kappa_\Lambda$, in order for the flow to approach as close as possible to the cuspy fixed point describing the critical behavior of the RFIM. (We achieve this by dichotomy.)\cite{footnote_UV}

\begin{figure}[tbp]
\centering
\includegraphics[width=.9 \linewidth]{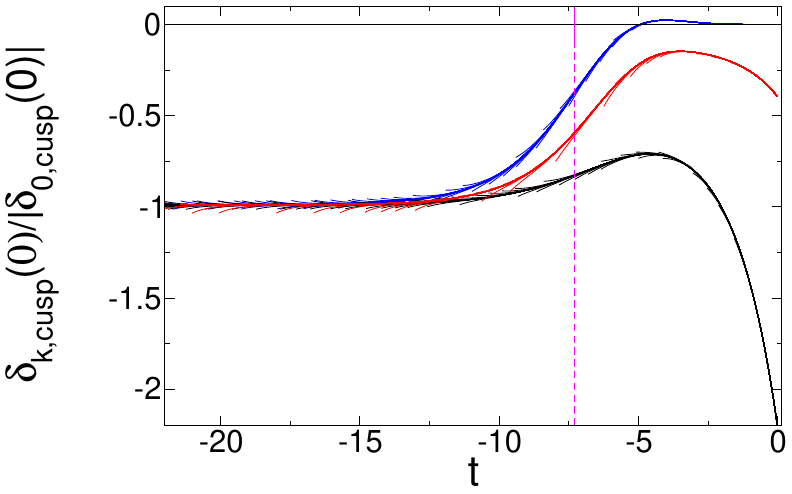}
\includegraphics[width=.9 \linewidth]{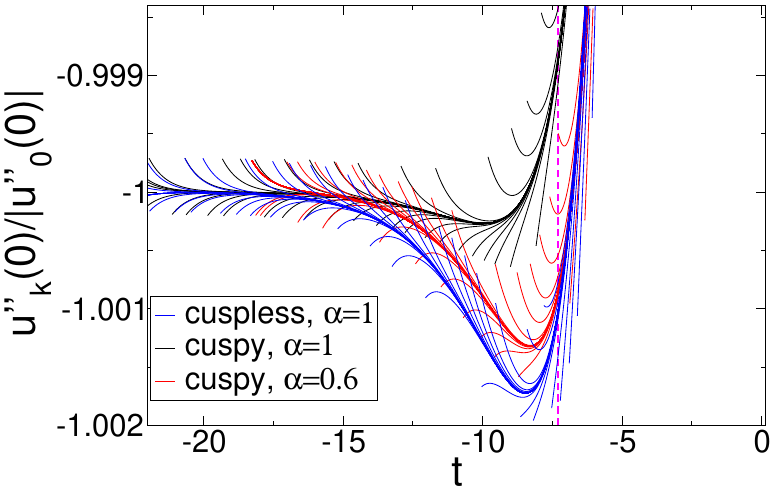}
\caption{NP-FRG flow of the amplitude of the cusp in the dimensionless second cumulant of the renormalized random field, $\delta_{k,{\rm cusp}}(\varphi=0)$, (top panel) and of the second derivative of the dimensionless potential $u''_k(\varphi=0)$ (bottom panel) as a function of the RG time $t=\ln(k/\Lambda)$, where $\Lambda$ is the UV scale, in $d=5$. The initial condition is fine-tuned (by dichotomy) to asymptotically approach the (critical) fixed point and $\delta_{k,{\rm cusp}}(\varphi=0)$ and $u''_k(\varphi=0)$ are divided by the modulus of their fixed-point value. Different settings are shown, one with the IR cutoff functions related by the SUSY-like form (Eq. (\ref{eq_SUSYregulator}) and a cuspless initial condition, one with the same IR cutoff functions ({\it i.e.}, $\alpha=1$) but a cuspy initial condition, and one with a modified relation between the cutoff functions with a prefactor $\alpha=0.6$ and a cuspy initial condition. The Larkin length (here, Larkin RG time $t_L$) is indicated by a vertical line. It becomes a crossover when the initial amplitude of the cusp departs from $0$. (However, due to numerical precision in the full functional numerical solution, there is a small nonzero value near the Larkin time even when starting from the cuspless condition, which also leads to a spurious rounding; the Larkin time can be crisply defined by expanding instead in one of the field arguments and detecting a divergence in the appropriate function, see the main text.) Note the scale of the vertical axis for $u''_k(\varphi=0)/\vert u''_0(\varphi=0)\vert$: To zoom in on the interesting region around and beyond the Larkin time, we do not display the early stages of the RG flow.}
\label{Fig_flow_cusp}
\end{figure}

We illustrate the results for $d=5$. We first show the NP-FRG flow of the amplitude of the cusp in the (dimensionless) second cumulant of the renormalized random field $\delta_{k,{\rm cusp}}(\varphi=0)$ in the top panel of Fig.~\ref{Fig_flow_cusp}. As anticipated, the Larkin length is not crisply defined for a nonzero initial cusp amplitude, but it remains as a noticeable crossover in the flow and its order of magnitude does not vary significantly with the amplitude of the cusp. (Note that the curve shown for the cuspless case is obtained from a full functional solution that requires discretizing all the fields and therefore, due to numerical precision, displays some systematic error in the vicinity of the Larkin time which leads to a spurious rounding; the Larkin scale can be accurately accessed by instead expanding the second cumulant in the difference between the two fields, which corresponds to the results shown in Fig.~\ref{Fig_Larkin_length} and to the vertical line in Fig.~\ref{Fig_flow_cusp}.) Quite notably, the flows corresponding to the 3 different settings only converge to their fixed-point value for RG times significantly beyond the Larkin one, {\it i.e.}, for length scales beyond the Larkin length. The IR cutoff $k$ plays a role similar to that of a finite system size $L$ in restricting the spatial extent of the fluctuations and, with a grain of salt, one can associate $k$ with $2\pi/L$.\cite{footnote_finite-size} This means that only with system sizes much larger than the Larkin length can one properly access the asymptotic critical behavior. We display in the bottom panel the second derivative of the dimensionless potential $u''_k(\varphi=0)$, {\it i.e.}, the squared mass. Again, the 3 curves corresponding to the different settings converge to their fixed-point value for times much beyond the Larkin time.

As we showed that when the dimension $d$ approaches $d_{DR}$ from below the Larkin length increases exponentially fast in $1/\sqrt{d_{DR}-d}$, proper access to the asymptotic critical behavior would require astronomically large system sizes in a simulation if the dimension happens to be very close to $d_{DR}$.

\section{Corrections to scaling in the vicinity of $d_{DR}$}
\label{sec:corrections}

To complement the discussion of the critical behavior of the RFIM in the vicinity of $d_{DR}$ we study the corrections to scaling. To do so, we consider the flows of specially tailored quantities\cite{delamotte_private} built from running renormalized observables, {\it e.g.}, the squared mass $u''_k(\varphi=0)$, the running anomalous dimension $\eta_k$ obtained from the field renormalization function $z_k(\varphi)$ (see Appendix~\ref{app:NP-FRG}), or the amplitude of the cusp $\delta_{cusp}(\varphi=0)$. For a generic running quantity $m_k\equiv m(t)$, with as before $t$ being the RG time, we define
\begin{equation}
\label{eq_f}
f_m(t)=\frac{d}{dt} \ln(\vert \frac{d}{dt} m(t)\vert)\,.
\end{equation}
In practice, a very small positive quantity $\sim 10^{-16}$ is added to the argument of the logarithm to avoid spurious singularities in the numerics. The flow of $f_m$ displays plateaus separated by transient regions. The values of the plateaus give a direct access to the exponents controlling the corrections to scaling, {\it i.e.}, some irrelevant eigenvalues around the fixed point, and, for very large RG time, to the relevant eigenvalue $1/\nu$ describing the escape from the fixed point due to imperfect numerical fine-tuning of the critical initial conditions at the UV scale [actually, the plateau for the relevant eigenvalue is at $-1/\nu$ due to the definition in Eq. (\ref{eq_f})].

\begin{figure}[tbp]
\centering
\includegraphics[width=.9 \linewidth]{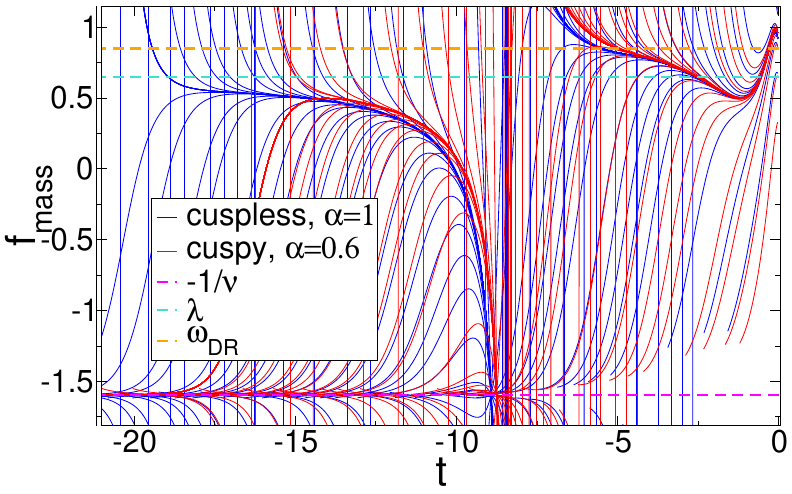}
\includegraphics[width=.9 \linewidth]{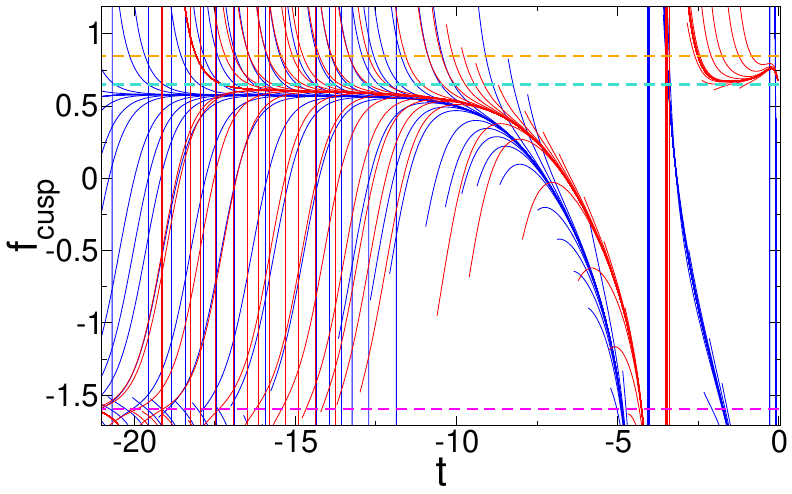}
\caption{Corrections to scaling for $d\lesssim d_{DR}$: NP-FRG flow of the quantities $f_{{\rm mass}}$, defined in Eq. (\ref{eq_f}))  from the (squared) mass $u''_k(\varphi=0)$, and $f_{{\rm cusp}}$ built from the amplitude of the cusp in the second cumulant  $\delta_{k,{\rm cusp}}(0)$, versus the RG time $t=\ln(k/\Lambda)$ in $d=5$.  We also show the asymptotic correction-to-scaling exponent $\omega=\lambda$, the dimensional-reduction correction-to-scaling exponent $\omega_{DR}$, and the relevant eigenvalue $-1/\nu$. The RG time $\vert t_L\vert $ associated with the Larkin length is about $7$ for $d=5$. The two different schemes, with cuspy or cuspless initial conditions and the appropriate NP-FRG equations, are shown. In all cases the initial conditions are fine-tuned by dichotomy to approach the critical fixed point.}
\label{Fig_flow_fm_before}
\end{figure}

We first discuss the case where $d<d_{DR}$.  We again consider the different settings, with different choices of IR cutoff functions associated with cuspy and cuspless initial conditions (see above). In Fig.~\ref{Fig_flow_fm_before}, we show the flow of $f_{{\rm mass}}$ and $f_{{\rm cusp}}$ for $d=5$.  (The behavior of the function associated with the running anomalous field dimension is very similar to that of $f_{{\rm mass}}$ and is not displayed here.). In $d=5$, we see first some ill-characterized behavior before and around the Larkin length, where no real plateau is found. (There seems to be an indication for $f_{{\rm mass}}$ that a short plateau around $\omega_{DR}\approx 0.85$ emerges; on the other hand the minimum of $f_{{\rm cusp}}$ at a value near $\lambda$ is accidental and is not found in other quantities or other dimensions.) This is followed by an approach to the fixed-point value that is governed by the correction-to-scaling exponent $\omega=\lambda\approx 0.65$ (more clearly reached for $f_{{\rm cusp}}$ than for $f_{{\rm mass}}$). Finally, because one cannot start exactly at the critical initial condition with an infinite precision, we also find a final regime governed by the relevant eigenvalue $-1/\nu\approx -1.59$ (with a different approach to the asymptote for $f_{{\rm mass}}$ and $f_{{\rm cusp}}$) as the flow is driven away from the vicinity of the fixed point. The situation is  similar in $d=5.1$ (not shown here), but the Larkin length is much larger.

The situation is quite different when $d$ is close but above $d_{DR}$, $d\gtrsim d_{DR}$. The critical theory at the fixed point now satisfies SUSY and the main scaling behavior is described by dimensional reduction. There are nonetheless small irrelevant eigenvalues, significantly smaller than $\omega_{DR}$ (see Fig.~\ref{Fig_eigenvalues}),  {\it but they are associated with eigenfunctions that depend on distinct copies (or replicas) and have no projection on the $1$-copy sector} ({\it i.e.}, when all field arguments are equal). To be more specific, we consider the second cumulant of the renormalized random field, which is generically of function of two replica fields: $\delta_k(\varphi_1,\varphi_2)$, or, after introducing the fields $\varphi=(\varphi_1+\varphi_2)/2$ and $y=(\varphi_2-\varphi_1)/2$, $\delta_k(\varphi,y)$. The function $\delta_k(\varphi,y)$ is even in $\varphi$, as a result of the statistical $Z_2$ symmetry of the model,  and even in $y$, as a result of replica permutation invariance. When the two replica fields become close, $y\to 0$, one can expand the function at the cuspless fixed point as
\begin{equation}
\label{eq_delta-expanded}
\delta_*(\varphi,y)=\delta_{*,0}(\varphi) +\frac 12 \delta_{*,2}(\varphi) y^2 + \cdots\,,
\end{equation}
where the equal-field ($y=0$) component satisfies the SUSY Ward identity, $\delta_{*,0}(\varphi)\equiv \delta_*(\varphi,\varphi)=z_*(\varphi)$ [see also Eq. (\ref{eq_SUSYward})]. In the vicinity of this fixed point, for small nonzero $k$, the second cumulant can be expanded as
\begin{equation}
\label{eq_delta-expanded}
\delta_k(\varphi,y)=\delta_{k,0}(\varphi) + \delta_{k,{\rm cusp}}(\varphi) \vert y\vert +\frac 12 \delta_{k,2}(\varphi) y^2 + O(\vert y\vert^3)\,,
\end{equation}
where we have now included nonanalytic terms in odd powers of $\vert y\vert$, which go to zero as $k\to 0$. The additional irrelevant eigenvalues that are not part of the $1$-replica dimensional-reduction spectrum are associated with $\delta_{k,{\rm cusp}}(\varphi)$ which flows to zero at the fixed point with the eigenvalue $\lambda$, with $\delta_{k,2}(\varphi)$ which flows to zero with the eigenvalue $\lambda_2$, etc.  

\begin{figure}[tbp]
\centering
\includegraphics[width=.9 \linewidth]{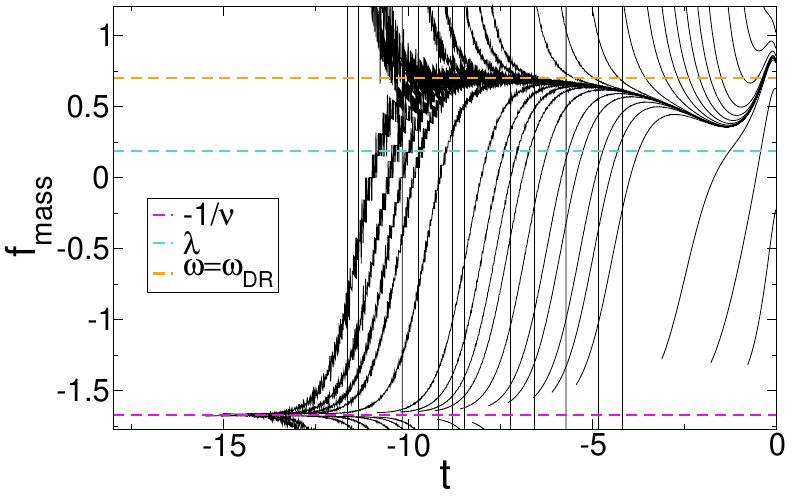}
\includegraphics[width=.9 \linewidth]{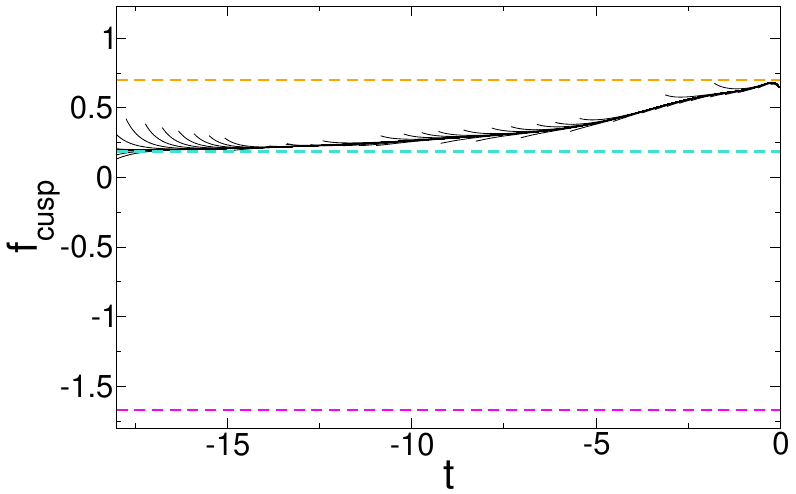}
\caption{Corrections to scaling: NP-FRG flow of the quantities [defined in Eq. (\ref{eq_f})] $f_{{\rm mass}}$, built from the squared mass $u''_k(\varphi=0)$, (top) and $f_{{\rm cusp}}$, built from the amplitude of the cusp in the second cumulant  $\delta_{k,{\rm cusp}}(0)$, (bottom) versus the RG time $t=\ln(k/\Lambda)$ in $d=5.17>d_{DR}$.  We also show in both cases the dimensional-reduction correction-to-scaling exponent $\omega_{DR}\approx 0.70$, the smallest irrelevant eigenvalue $\lambda\approx 0.19$ associated with a cuspy eigenfunction in the 2-copy sector, and the dimensional-reduction relevant eigenvalue $-1/\nu\approx -1.67$. The initial conditions are fine-tuned by dichotomy to approach the critical fixed point, at which SUSY is satisfied; they are chosen with a cusp in the second cumulant For $f_{{\rm mass}}$, due to imperfect numerical fine-tuning of the critical initial conditions, the long-time RG time behavior displays a final  plateau associated with relevant eigenvalue $1/\nu$ shown as a dashed line. The behavior of  $f_{{\rm cusp}}$ at long RG times is different. Note that the flow of $f_{{\rm mass}}$ is only sensitive to $\omega_{DR}$, contrary to that of $f_{{\rm cusp}}$ which is only sensitive to $\lambda$.}
\label{Fig_flow_fm_after}
\end{figure}

We stress again that the small additional irrelevant eigenvalues $\lambda$, $\lambda_2$, etc., are only visible in the corrections to scaling of observables probing correlations between {\it distinct} copies of the system. We illustrate this point by showing the NP-FRG flows of the two quantities $f_{{\rm mass}}$ and $f_{{\rm cusp}}$ introduced above in Eq. (\ref{eq_f}) in Fig.~\ref{Fig_flow_fm_after} for the dimension $d=5.17$ and we consider a cuspy initial condition (otherwise $\delta_{k,{\rm cusp}}$ and $f_{{\rm cusp}}$ stay identically zero), but with IR cutoff functions that satisfy the SUSY Ward identity so that the proper fixed point can be reached. One can see that the flow of the quantity $f_{{\rm mass}}$ associated with the squared mass is only sensitive to the dimensional-reduction correction-to-scaling exponent $\omega_{DR}\approx 0.70$ and, at very long RG time where the system finally escapes from the vicinity of the fixed point, to the relevant eigenvalue $-1/\nu \approx -1.67$. (The same is true for the quantity associated with the field renormalization, $f_\eta$ but it is not displayed here.) On the other hand, the correction to scaling of the amplitude of the cusp, which is a genuine characteristics of the $2$-copy sector, is  controlled by the small eigenvalue $\lambda\approx 0.19$, as seen in the flow of $f_{{\rm cusp}}$. At very long RG time $f_{{\rm cusp}}$ does not follow the behavior of $f_{{\rm mass}}$ and does not  seem to converge to the relevant eigenvalue $-1/\nu$; this indicates that the projection of the eigenvector associated with $-1/\nu$ on the cusp amplitude in the second cumulant vanishes.

\section{Conclusion}
\label{sec:conclusion}

We have shown that the NP-FRG prediction for the RFIM of the critical dimension $d_{DR}\approx5.1$ separating a region ($d>d_{DR}$) where the critical behavior follows dimensional reduction (except in irrelevant multi-copy cuspy directions\cite{FPbalog,tarjus13,balog_activated,tissier_pertFRG}) and is controlled by a SUSY fixed point from a region ($d<d_{DR}$) where both dimensional reduction and SUSY are broken at criticality is compatible with the recent large-scale numerical simulations in $d=5$.\cite{fytas17,fytas19} In particular, the peculiar mechanism through which the SUSY-broken fixed point emerges from the SUSY fixed point at $d_{DR}$ explains why correction to scaling can be controlled by a rather large exponent $\omega$ despite the fact that $d=5$ is not far from $d_{DR}$. 

We have also discussed the properties of the corrections to scaling in the immediate vicinity of $d_{DR}$ and their consequences for finite-size lattice simulations. Indeed, the NP-FRG prediction for the critical dimension, $d_{DR}\approx 5.1$, is an approximate result only. (Work is now in progress to assess the uncertainty in its determination.) We have argued that for a dimension very close but below $d_{DR}$ the existence of a very large Larkin length that diverges exponentially fast as one approaches $d_{DR}$ precludes any detection of the asymptotic large-scale behavior in simulations with presently achievable system sizes.  On the other hand for a dimension very close but above $d_{DR}$, the usual observables considered in finite-size analyses, masses or correlation lengths, main critical exponents, etc., which are all properties of the $1$-copy sector, {\it i.e.}, quantities evaluated for equal replica fields, are only affected by the dimensional-reduction correction-to-scaling exponent $\omega_{DR}$, which is large near $d_{DR}$; it is only by studying quantities that are genuinely associated with $2$ or more distinct copies that one would capture the small irrelevant eigenvalues associated with singular eigenfunctions.

If the correction to scaling in $d=5$ turns out not to be the ``smoking gun'' proposed in Ref.~[\onlinecite{fytas19}], can one propose a crisper test of our prediction of a critical $d_{DR}$ that could be checked in computer simulations of the RFIM? It seems that the $1$-d long-range RFIM model, in which the interactions decay as a power law with distance, $r^{-(d+\sigma)}$ but the correlations of the bare random field are short-ranged,\cite{bray_LR1D,balog_LR1D} could be a good candidate. Dimension $d$ is fixed but in some sense varying the range exponent $\sigma$ has a similar effect to that of changing $d$ in the short-range RFIM: for $\sigma \leq 1/3$ the critical behavior is controlled by a Gaussian fixed point whereas no phase transition takes place for $\sigma \geq 1/2$; between the two values a critical point with nontrivial $\sigma$-dependent exponents exists. Our prediction based on the NP-FRG is that a critical value $\sigma_c\approx 0.38$ separates a regime where the fixed point is ``cuspless''  from a regime where it is ``cuspy'', with the same boundary-layer mechanism governing the appearance of the cuspy fixed point for $\sigma \to \sigma_c^+$.\cite{balog_LR1D} The discontinuity in the eigenvalue $\lambda$ is now large in $\sigma_c$ (it goes from $0.7$ to $0$) and the change of behavior, including the correction to scaling, should be more readily detectable in a lattice simulation.

\begin{acknowledgments}
IB acknowledges the support of the Croatian Science Foundation Project No. IP-2016-6-3347 and the QuantiXLie Centre of Excellence, a project cofinanced by the Croatian Government and European Union through the European Regional Development Fund - the Competitiveness and Cohesion Operational Programme (Grant KK.01.1.1.01.0004).
\end{acknowledgments}

\appendix

\section{Nonperturbative FRG}
\label{app:NP-FRG}

We summarize here the main features of the NP-FRG description of the equilibrium critical behavior of RFIM developed in Refs.~[\onlinecite{tarjus04,tissier06,tissier11,tarjus20}]. The central quantity is the so-called ``effective average action'' $\Gamma_k$,\cite{wetterich93,berges02} in which only fluctuations of modes with momentum larger than an infrared cutoff $k$ are effectively taken into account.  In the language of  magnetic systems, $\Gamma_k$ is the Gibbs free-energy functional of the local order parameter field obtained after a coarse-graining down to the (momentum) scale $k$. The effective average action obeys an exact RG equation under the variation of the infrared cutoff $k$.\cite{wetterich93}

In the presence of disorder, the generating (free-energy) functionals are sample-dependent, \textit{i.e.} random, and should therefore be characterized either by their probability distribution or by their cumulants. The latter description is more convenient as it focuses on quantities---cumulants and associated Green's functions---which are translationally invariant and can be generated through the introduction of copies (or ``replicas'') of the original system that are submitted to distinct external sources.\cite{tarjus04,tissier11,tarjus20} The effective average action that generates the cumulants of the renormalized disorder, $\Gamma_k[\{\phi_a\}]$, then depends on the local order parameter fields associated with the various copies $a$. It satisfies the following exact FRG flow equation:\cite{tarjus04,tissier11}
\begin{equation}
\begin{aligned}
\label{eq_erg}
&\partial_k\Gamma_k\left[\{ \phi_a\}\right ]=\\&
\dfrac{1}{2} \int \frac{d^{d} q}{(2\pi)^d} \sum_{ab}  \partial_k R_{k}^{ab}(q^2) \bigg (\left[ \Gamma _k^{(2)}+ R_k\right]^{-1}\bigg)_{q,-  q}^{ab},
\end{aligned}
\end{equation}
where $\Gamma_k^{(2)}$ is the matrix formed by the second functional derivatives of $\Gamma_k$ with respect to the fields $\phi_a( q)$ and  $R_{k}^{ab}(q^2)=\widehat{R}_k(q^2) \delta_{ab}+ \widetilde{R}_k(q^2)$, where $\widehat{R}_k$ and $\widetilde{R}_k$ are infrared cutoff functions that enforce the decoupling of the low- and high-momentum modes at the scale $k$ and reduce the fluctuations induced by the disorder: More precisely, the function $\widehat{R}_k(q^2)$ adds a mass $\sim k^{2-\eta}$ (where $\eta$ is a running anomalous dimension of the fields) to modes with $q^2\lesssim k^2$ and decays rapidly to zero for $q^2 \gtrsim k^2$, whereas the function $\widetilde{R}_k(q^2)$ reduces the fluctuations of the bare random field. The formalism can be upgraded to a superfield theory in order to describe the physics directly at zero temperature (\textit{i.e.}, the ground-state properties), which allows one to make the underlying SUSY\cite{parisi79} explicit and describe its spontaneous breaking.\cite{tissier11}

From Eq.~(\ref{eq_erg}), one can derive a hierarchy of coupled RG flow equations for the cumulants of the renormalized disorder, $\Gamma_{k1}[\phi_1]$, $\Gamma_{k2}[\phi_1,\phi_2]$, etc., that are obtained from $\Gamma_k[\{\phi_a\}]$ through an expansion in increasing number of unrestricted  sums over copies, $\Gamma_k\left[\{ \phi_a\}\right ]=\sum_a \Gamma_{k1}[\phi_a]-\frac{1}{2}\sum_{a,b}\Gamma_{k2}[\phi_a,\phi_b] + \cdots$. Alternatively, one can consider the exact hierarchy of RG equations for the field derivatives of these quantities, $\Gamma_{k1,x}^{(1)}[\phi_1]$, $\Gamma_{k2,xy}^{(1,1)}[\phi_1,\phi_2]$, etc., which represents the cumulants of the renormalized random field and are the objects that are naturally obtained from the superfield construction.

At the UV scale, say $k=\Lambda$, the effective average action reduces to some ``bare'' action of the multy-copy system, which generates the cumulants of the renormalized disorder at the mean-field level. At the end of the flow, when $k=0$, all the fluctuations are incorporated and one recovers the full effective action $\Gamma[\{ \phi_a\} ]$ which is the generating functional of the $1$-particle irreducible ($1$-PI) correlation functions that are associated with the  cumulants of the fully renormalized disorder. The minimal truncation that  already contains the key features for a nonperturbative study of the critical physics of the RFIM is the following:
\begin{equation}
\label{eq_ansatz_Gamma_k}
\begin{aligned}
&\Gamma_{k1,x}^{(1)}[\phi]=  U'_{k}( \phi_1(x)) + \frac 12 [\frac{\delta}{\delta \phi_1(x)}Z_{k}(\phi_1(x))  [\partial  \phi_1(x)]^2 \\&
\Gamma_{k2,x_1 x_2}^{(1,1)}[\phi_1,\phi_2]=\delta^{(d)}(x_1-x_2) \Delta_{k}( \phi_1(x_1),  \phi_2(x_1)) \\&
\Gamma_{kp,x_1\cdots x_p}^{(1,\cdots,1)}[\phi_1,\cdots,\phi_p]=0\; {\rm for}\; p\geq 3\,,
\end{aligned}
\end{equation}
with three functions, the field renormalization function $Z_k(\phi)$, the potential $U_k(\phi)$ [or its derivative $U'_k(\phi)$], and the second cumulant of the renormalized random field $\Delta_k(\phi_1,\phi_2)$, to be determined. On the other hand, to ensure that there is no explicit breaking of SUSY the IR cutoff functions must satisfy the relation
\begin{equation}
\widetilde{R}_k(q^2)=-(\Delta_k/Z_k)\frac{\partial}{\partial q^2}\widehat{R}_k(q^2),
\end{equation}
with $\Delta_k$ the strength of the renormalized random field and $Z_k$ the field renormalization constant [defined, {\it e.g.}, by $\Delta_k=\Delta_k(\phi_1=0,\phi_2=0)$ and $Z_k=Z_k(\phi=0)$]. Inserting the above ansatz in the exact RG equations for the cumulants leads to a set of coupled flow equations for the three functions $Z_{k}(\phi)$, $U'_k(\phi)$ and $\Delta_{k}( \phi_1,  \phi_2)$. The RG is functional as its central objects are functions instead of coupling constants.

To cast the NP-FRG flow equations in a form that is suitable for searching for zero-temperature fixed points describing the critical behavior of the RFIM,\cite{villain84,fisher86} one has to introduce appropriate scaling dimensions.  This entails defining a renormalized temperature $T_k$ which flows to zero as $k \rightarrow 0$. Near such a  fixed point, one has the following scaling dimensions:  $T_k \sim k^\theta$,  $Z_{k} \sim k^{-\eta}$, $\phi_a  \sim k^{(d-4+\bar \eta)/2}$, with $\theta$ and $\bar \eta$ related through $\theta=2+\eta-\bar \eta$, as well as $U'_k\sim k^{d-\theta-(d-4+\bar \eta)/2}$ and  $\Delta_k \sim k^{-(2\eta- \bar \eta)}$. Letting the dimensionless counterparts of $U_k$, $Z_k$, $\Delta _k$,  $\phi$  be denoted by lower-case letters, $u_k, z_k,\delta _k,  \varphi$,  the resulting flow equations can be symbolically written as
\begin{equation}
\label{eq_flow_dimensionless}
\begin{split}
&\partial_t u'_k(\varphi)=\beta_{u'}(\varphi),\\&
\partial_t z_k(\varphi)=\beta_{z}(\varphi),\\&
\partial_t \delta_k(\varphi_1,\varphi_2)=\beta_{\delta}(\varphi_1,\varphi_2),
\end{split}
\end{equation}
where $t=\log(k/\Lambda)$. The beta functions themselves depend on the functions $u_k'$, $z_k$, $\delta_k$ and their derivatives [in addition, the running anomalous dimensions $\eta_k$ and $\bar\eta_k$ are fixed by the conditions $z_k(0)=\delta_k(0,0)=1$]. Their expressions are given in Ref.~[\onlinecite{tissier11}] for the zero (bare) temperature case.

The beta functions also depend on the dimensionless cutoff functions, which are defined from $\widehat R_k(q^2)=Z_k k^2 \,\widehat r(q^2/k^2)$ and $\,\widetilde R_k(q^2)=\Delta_k \widetilde r(q^2/k^2)$ with $\widetilde r(x^2)=-\partial_{x^2}\widehat r(x^2)$, where we chose for the function $\widehat r(x^2)$ the form $(a+b x^2+c x^4)e^{-x^2}$. The parameters $a$, $b$ and $c$ of the function can be further optimized by invoking the principle of minimum sensitivity\cite{PMS} and we find good stability of the results for $a\approx1.7$, $b\approx 0.81$ and $c\approx 0.14$. (The value of the critical dimension $d_{DR}$ does not significantly depend on the values of the parameters.)

In this work, we study either the fixed-point solution of Eqs. (\ref{eq_flow_dimensionless}) (obtained by setting the left-hand sides to zero) and the spectrum of eigenvalues obtained from solving the eigenvalue equations derived by linearizing the beta functions around the fixed point or the flow of the various functions as a function of the RG time $t$ for initial conditions that are fine-tuned by dichotomy to approach as close as possible the fixed at large RG time. For the numerical resolution the fields are discretized on a grid.

\section{Toy model of NP-FRG flow equations around $d_{DR}$}
\label{app:toy}

We consider a toy model inspired by the 1-loop flow equation of the RF$O(N)$M, which we introduced in Ref.~[\onlinecite{FPbalog}] to investigate in detail the scenario of appearance and disappearance of fixed points in the whole plane $(N,d)$. It consists in a partial differential equation which mimics the RG flow of the second cumulant of the renormalized random field, represented here by a function $\Delta(z)$ where $z\in[-1,1]$:
\begin{equation}
  \label{eq:betatoy}
  \begin{split}
  \partial_t \Delta_k(z)=&\Delta_k(z)-\Delta_k(z)\Delta'_k(z)-(\Delta_k(z)-z \Delta_k'(z)) \Delta_k(1)
  \\&
  +B\big [ \Delta_k(1)-z \Delta_k(z)\big ]\big [\Delta_k(z)+z \Delta_k'(z)\big ] +
  \\&
  \frac A2 (1-z^2)\Delta_k'(z)\big [2z\Delta_k'(z)-(1-z^2)\Delta_k''(z)\big ] \,.
  \end{split}
\end{equation}
The beta function depends on two parameters, $A$ and $B$, which replace the two parameters $d$ and $N$ of the RF$O(N)$M.

We showed that for $A>3/2$ and $B<1/20$, the critical ({\it i.e.}, once unstable) cuspy fixed point emerges from the collapse of two cuspless fixed points when $B_{DR}(A)=1/[8(1+A)]$ via a boundary-layer mechanism, in a manner that appears very similar to the appearance of the cuspy critical fixed point in the RFIM below $d_{DR}\approx 5.1$. More specifically, one can show that within a boundary layer of width $\epsilon$, where $\epsilon=B-B_{DR}(A)$, the function $\Delta_k(z)$ can be written as
\begin{equation}
  \label{eq:def_f}
  \Delta_k(z)=1-\epsilon  f\left(\sqrt{\frac{1-z}\epsilon}\right) \,, 
\end{equation}
where $\epsilon \to 0$ and $f(y)$ has a cusp at small $y=\sqrt{1-z}/\epsilon$, $f(y)=f(0)+ay+f_2(y^2/2)$, and at large $y$ converges to the outer solution that has a regular expansion in $1-z$, $f(y)\sim b y^2 + c$, when $y\to \infty$. We find $f(0)=-a^2(2A+3)^2/[16(A+1)]$, $f_2=2(8A+5)/[3(8A^2+16A+9)]$, $b=1/(4A+3)$, and $c=-(8A+3)/3$. The outer solution in the region where $\epsilon \ll 1-z \ll1$ can be expressed at the fixed point in leading order in $(1-z)$ and $\epsilon$ as 
\begin{equation}
  \label{eq:outer}
\Delta_*(z)=1+ (1-z)[-\frac 1{4A+3} +\sqrt \epsilon\, d] +\epsilon f(0)\frac{8A+3}{3})+\cdots 
\end{equation}
where  $d$ has an (unilluminating) expression in terms of $A$ and $f(0)$ and where we have used constraints coming from matching the inner and outer solutions.

To obtain the lowest irrelevant eigenvalue $\lambda$ when $B\geq B_{DR}(A)$ (which corresponds to $d\leq d_{DR}$) we rewrite the cumulant function as $ \Delta_k(z)=\Delta_*(z)+k^\lambda \delta(z)$ and we linearize the flow equation, Eq.~(\ref{eq:betatoy}), around the fixed point. Both in the inner and in the outer region, this leads to $\lambda=0$ when $B=B_{DR}(A)$ and, by considering the linearized equation for the outer solution in the region where $\epsilon \ll 1-z \ll1$, we immediately obtain that $\lambda \propto \sqrt \epsilon$, {\it i.e.}, $\lambda \propto \sqrt{B-B_{DR}(A)}$ for a given $A>3/2$ (or a similar relation with $A-A_{DR}$ at $B<1/20$ fixed). 

The irrelevant eigenvalue $\lambda$ associated with a cuspy perturbation around the cuspless fixed point therefore has a discontinuity in $B_{DR}$,\cite{FPbalog}
\begin{equation}
\lambda(B_{DR}^-)-\lambda(B_{DR}^+)=\lambda(B_{DR}^-)=\frac{2A-3}{4(4A+3)}=\frac{1-20B_{DR}}{8(1-2B_{DR})}
\end{equation}
with $A>3/2$ (correspondingly, $B_{DR}<1/20$), and, from the derivation above and the result in [\onlinecite{FPbalog}], it has a square-root behavior on both sides of $B_{DR}$. In addition, it is easily checked (see Eq. (20) in Ref. [\onlinecite{FPbalog}]) that for $B<B_{DR}$ there is an irrelevant eigenvalue $\lambda_2$, which is associated with a perturbation that has no cusp and which vanishes at $B_{DR}$ with a square-root singularity. 



We can also show in this toy model that the Larkin length grows exponentially as one approaches the critical value $B_{DR}(A)$ (or $A_{DR}(B)$ from the cuspy phase ($B>B_{DR}$). We find that 
\begin{equation}
\vert t_L\vert \sim \frac 1{\sqrt{B-B_{DR}}}
\end{equation}
for $A>3/2$ fixed, or a similar relation with $A-A_{DR}$ for $B<1/20$ fixed. The Larkin length can be obtained from the RG time at which the first derivative of the function $\Delta_k(z)$ in $z=1$ diverges when starting from cuspless initial conditions and fixing $\Delta_k(1)=1$ which corresponds to the cuspless fixed point. The flow equation for $\Delta'_k(1)$ is then easily obtained as\cite{FPbalog} 
\begin{equation}
\begin{aligned}
  \partial_t \Delta_k'(1)=&-B\big [1+\Delta_k'(1)\big ]^2\\&
  +\Delta_k'(1)\big [1-(1+2A)\Delta_k'(1)\big ] \,.
\end{aligned}
\end{equation}
In the region where the true fixed point has a cusp, {\it i.e.}, for $B>B_{DR}(A)$ with $A<3/2$, the appearance of the cusp in $\sqrt{1-z}$ in $\Delta_k(z)$ is  signaled by the divergence of $\Delta'_k(1)$. From solving the above equation, one finds that this takes place when $\vert t_L\vert=\pi/2 + \arctan[(1-2B)/\sqrt{8B(1+A)-1}]$. In the vicinity of $B_{DR}(A)=1/[8(1+A)]$, this immediately leads to $\vert t_L\vert \sim 1/\sqrt{B-B_{DR}(A)}$, as announced.


\begin{thebibliography}{15}
\bibitem{footnote_BL}
This boundary layer mechanism should not be confused with the thermal boundary layer that is present the approach to the zero-temperature fixed point of the RFIM\cite{tarjus04,tissier06,tissier11,balog_activated} (or of an elastic manifold in a random environment as well\cite{chauve_creep,balents-doussal_BL}) when the bare temperature is different from zero and describes the rounding by temperature of the cusps in the functional dependence of the cumulants of the renormalized random field.
\bibitem{footnote_omega}
Note that in the paper on SUSY in $d=5$,\cite{fytas19} the authors do not use their previously determined value of $\omega$ but take the dimensional-reduction prediction $\omega_{DR}\approx 0.83$ to check if SUSY is broken or not.
\bibitem{footnote_fixed-points}
When the IR cutoff $k\to 0$, the effective average action as a functional of the dimensionful fields and all physical quantities, such as the critical exponents, become independent of the IR cutoff functions. The fixed point itself on the other hand, which is characterized by the effective action as a functional of dimensionless fields, generically depends on the choice of the (dimensionless) IR cutoff functions. Scale invariance that applies at the critical point relates the variation of the fields and of the coordinates with the variation of the cutoff functions when $k$ is infinitesimal but not exactly zero. As a result different cutoff functions lead to different fixed points that nonetheless describe the very same long-distance physics.
\bibitem{footnote_UV}
Note that the UV scale $k_{UV}=\Lambda$ is now no longer directly related to the inverse of a lattice spacing, $2\pi/a$,  and, according to Ref.~[\onlinecite{machado10}], should be envisaged as much bigger.
\bibitem{footnote_finite-size}
The length $L$ is measured is units of the lattice spacing $a$ whereas the IR cutoff $k$ is measured in units of the UV scale $\Lambda$. As discussed in the footnote [\onlinecite{footnote_UV}],  $\Lambda$ in a field-theoretical setting should be thought as much bigger than $2\pi/a$,  which should be accounted for when comparing $L$ and $2\pi/k$.
\end{thebibliography}
\end{document}